\documentclass[twocolumn]{aastex631_arxiv}
\usepackage{amsmath}
\usepackage{graphics}
\usepackage{graphicx}
\usepackage{amssymb}
\usepackage{footmisc}

\begin{document}

\title{Estimating Electron Densities in the Middle Solar Corona using White-light and Radio Observations}
\correspondingauthor{Surajit Mondal}
\email{surajit@ncra.tifr.res.in}

\author[0000-0002-2325-5298]{Surajit Mondal}
\affiliation{Center for Solar-Terrestrial Research, New Jersey Institute of Technology, \\
323 M L King Jr Boulevard, Newark, NJ 07102-1982, USA}

\author[0000-0002-3089-3431]{Shaheda Begum Shaik}
\affiliation{George Mason University, Fairfax, VA, USA}
\affiliation{U.S.\ Naval Research Laboratory, Washington, DC, USA}

\author[0000-0001-9027-8249]{Russell A. Howard}
\affiliation{Johns Hopkins University Applied Physics Laboratory, Laurel, MD, USA}

\author[0000-0001-6855-5799]{Peijin Zhang}
\affiliation{Center for Solar-Terrestrial Research, New Jersey Institute of Technology, \\
323 M L King Jr Boulevard, Newark, NJ 07102-1982, USA}
\affiliation{Cooperative Programs for the Advancement of Earth System Science, University Corporation for Atmospheric Research, Boulder, CO, USA}

\author[0000-0002-0660-3350]{Bin Chen}
\affiliation{Center for Solar-Terrestrial Research, New Jersey Institute of Technology, \\
323 M L King Jr Boulevard, Newark, NJ 07102-1982, USA}

\author[0000-0002-1810-6706]{Xingyao Chen}
\affiliation{Center for Solar-Terrestrial Research, New Jersey Institute of Technology, \\
323 M L King Jr Boulevard, Newark, NJ 07102-1982, USA}

\author[0000-0003-2872-2614]{Sijie Yu}
\affiliation{Center for Solar-Terrestrial Research, New Jersey Institute of Technology, \\
323 M L King Jr Boulevard, Newark, NJ 07102-1982, USA}

\author{Dale Gary}
\affiliation{Center for Solar-Terrestrial Research, New Jersey Institute of Technology, \\
323 M L King Jr Boulevard, Newark, NJ 07102-1982, USA}

\author{Marin M. Anderson}
\affiliation{Owens Valley Radio Observatory, California Institute of Technology, Big Pine, CA 93513, USA}
\affiliation{Jet Propulsion Laboratory, California Institute of Technology, Pasadena, CA 91011, USA}
\author{Judd D. Bowman}
\affiliation{School of Earth and Space Exploration, Arizona State University, Tempe, AZ 85287, USA}
\author{Ruby Byrne}
\affiliation{Cahill Center for Astronomy and Astrophysics, California Institute of Technology, Pasadena, CA 91125, USA}
\affiliation{Owens Valley Radio Observatory, California Institute of Technology, Big Pine, CA 93513, USA}
\author{Morgan Catha}
\affiliation{Owens Valley Radio Observatory, California Institute of Technology, Big Pine, CA 93513, USA}
\author{Sherry Chhabra}
\affiliation{Center for Solar-Terrestrial Research, New Jersey Institute of Technology, \\
323 M L King Jr Boulevard, Newark, NJ 07102-1982, USA}
\affiliation{George Mason University, Fairfax, VA 22030, USA}
\author{Larry D'Addario}
\affiliation{Cahill Center for Astronomy and Astrophysics, California Institute of Technology, Pasadena, CA 91125, USA}
\affiliation{Owens Valley Radio Observatory, California Institute of Technology, Big Pine, CA 93513, USA}
\author{Ivey Davis}
\affiliation{Cahill Center for Astronomy and Astrophysics, California Institute of Technology, Pasadena, CA 91125, USA}
\affiliation{Owens Valley Radio Observatory, California Institute of Technology, Big Pine, CA 93513, USA}
\author{Jayce Dowell}
\affiliation{University of New Mexico, Albuquerque, NM 87131, USA}
\author{Gregg Hallinan}
\affiliation{Cahill Center for Astronomy and Astrophysics, California Institute of Technology, Pasadena, CA 91125, USA}
\affiliation{Owens Valley Radio Observatory, California Institute of Technology, Big Pine, CA 93513, USA}
\author{Charlie Harnach}
\affiliation{Owens Valley Radio Observatory, California Institute of Technology, Big Pine, CA 93513, USA}
\author{Greg Hellbourg}
\affiliation{Cahill Center for Astronomy and Astrophysics, California Institute of Technology, Pasadena, CA 91125, USA}
\affiliation{Owens Valley Radio Observatory, California Institute of Technology, Big Pine, CA 93513, USA}
\author{Jack Hickish}
\affiliation{Real-Time Radio Systems Ltd, Bournemouth, Dorset BH6 3LU, UK}
\author{Rick Hobbs}
\affiliation{Owens Valley Radio Observatory, California Institute of Technology, Big Pine, CA 93513, USA}
\author{David Hodge}
\affiliation{Cahill Center for Astronomy and Astrophysics, California Institute of Technology, Pasadena, CA 91125, USA}
\author{Mark Hodges}
\affiliation{Owens Valley Radio Observatory, California Institute of Technology, Big Pine, CA 93513, USA}
\author{Yuping Huang}
\affiliation{Cahill Center for Astronomy and Astrophysics, California Institute of Technology, Pasadena, CA 91125, USA}
\affiliation{Owens Valley Radio Observatory, California Institute of Technology, Big Pine, CA 93513, USA}
\author{Andrea Isella}
\affiliation{Department of Physics and Astronomy, Rice University, Houston, TX 77005, USA}
\author{Daniel C. Jacobs}
\affiliation{School of Earth and Space Exploration, Arizona State University, Tempe, AZ 85287, USA}
\affiliation{Rice Space Institute, Rice University, Houston, TX 77005, USA}
\author{Ghislain Kemby}
\affiliation{Owens Valley Radio Observatory, California Institute of Technology, Big Pine, CA 93513, USA}
\author{John T. Klinefelter}
\affiliation{Owens Valley Radio Observatory, California Institute of Technology, Big Pine, CA 93513, USA}
\author{Matthew Kolopanis}
\affiliation{School of Earth and Space Exploration, Arizona State University, Tempe, AZ 85287, USA}
\author{Nikita Kosogorov}
\affiliation{Cahill Center for Astronomy and Astrophysics, California Institute of Technology, Pasadena, CA 91125, USA}
\affiliation{Owens Valley Radio Observatory, California Institute of Technology, Big Pine, CA 93513, USA}
\author{James Lamb}
\affiliation{Owens Valley Radio Observatory, California Institute of Technology, Big Pine, CA 93513, USA}
\author{Casey Law}
\affiliation{Cahill Center for Astronomy and Astrophysics, California Institute of Technology, Pasadena, CA 91125, USA}
\affiliation{Owens Valley Radio Observatory, California Institute of Technology, Big Pine, CA 93513, USA}
\author{Nivedita Mahesh}
\affiliation{Cahill Center for Astronomy and Astrophysics, California Institute of Technology, Pasadena, CA 91125, USA}
\affiliation{Owens Valley Radio Observatory, California Institute of Technology, Big Pine, CA 93513, USA}
\author{Brian O'Donnell}
\affiliation{Center for Solar-Terrestrial Research, New Jersey Institute of Technology, \\
323 M L King Jr Boulevard, Newark, NJ 07102-1982, USA}
\author{Kathryn Plant}
\affiliation{Owens Valley Radio Observatory, California Institute of Technology, Big Pine, CA 93513, USA}
\affiliation{Jet Propulsion Laboratory, California Institute of Technology, Pasadena, CA 91011, USA}
\author{Corey Posner}
\affiliation{Owens Valley Radio Observatory, California Institute of Technology, Big Pine, CA 93513, USA}
\author{Travis Powell}
\affiliation{Owens Valley Radio Observatory, California Institute of Technology, Big Pine, CA 93513, USA}
\author{Vinand Prayag}
\affiliation{Owens Valley Radio Observatory, California Institute of Technology, Big Pine, CA 93513, USA}
\author{Andres Rizo}
\affiliation{Owens Valley Radio Observatory, California Institute of Technology, Big Pine, CA 93513, USA}
\author{Andrew Romero-Wolf}
\affiliation{Jet Propulsion Laboratory, California Institute of Technology, Pasadena, CA 91011, USA}
\author{Jun Shi}
\affiliation{Cahill Center for Astronomy and Astrophysics, California Institute of Technology, Pasadena, CA 91125, USA}
\author{Greg Taylor}
\affiliation{University of New Mexico, Albuquerque, NM 87131, USA}
\author{Jordan Trim}
\affiliation{Owens Valley Radio Observatory, California Institute of Technology, Big Pine, CA 93513, USA}
\author{Mike Virgin}
\affiliation{Owens Valley Radio Observatory, California Institute of Technology, Big Pine, CA 93513, USA}
\author{Akshatha Vydula}
\affiliation{School of Earth and Space Exploration, Arizona State University, Tempe, AZ 85287, USA}
\author{Sandy Weinreb}
\affiliation{Cahill Center for Astronomy and Astrophysics, California Institute of Technology, Pasadena, CA 91125, USA}
\author{Scott White}
\affiliation{Owens Valley Radio Observatory, California Institute of Technology, Big Pine, CA 93513, USA}
\author{David Woody}
\affiliation{Owens Valley Radio Observatory, California Institute of Technology, Big Pine, CA 93513, USA}
\author{Thomas Zentmeyer}
\affiliation{Owens Valley Radio Observatory, California Institute of Technology, Big Pine, CA 93513, USA}

\begin{abstract}

The electron density of the solar corona is a fundamental parameter in many areas of solar physics. Traditionally, routine estimates of coronal density have relied exclusively on white-light observations. However, these density estimates, obtained by inverting the white-light data, require simplifying assumptions, which may affect the robustness of the measurements. Hence, to improve the reliability of coronal density measurements, it is highly desirable to explore other complementary methods. In this study, we estimate the coronal electron densities in the middle corona, between approximately $1.7-3.5R_\odot$, using low-frequency radio observations from the recently commissioned Long Wavelength Array at the Owens Valley Radio Observatory (OVRO-LWA). The results demonstrate consistency with those derived from white-light coronagraph data and predictions from theoretical models. We also derive a density model valid between 1.7--3.5 $r_\odot$ and is given by $\rho (r')=1.27r'^{-2}+29.02r'^{-4}+71.18r'^{-6}$, where $r'=r/R_\odot$, and $r$ is the heliocentric distance. OVRO-LWA is a solar-dedicated radio interferometer that provides science-ready images with low latency, making it well-suited for generating regular and independent estimates of coronal densities to complement existing white-light techniques.

\end{abstract}

\section{Introduction}

Electron density is a key physical parameter necessary for describing the solar corona, and, hence, it is routinely used in many different branches of solar physics, including seemingly disparate and unrelated fields such as flare studies and solar wind acceleration. There is a host of different techniques to measure the coronal electron density. One widely used technique is differential emission measure (DEM) analysis, which is used to determine the density and temperature of multi-thermal coronal plasma. It involves simultaneously modeling data from multiple broadband, temperature-sensitive filters. While this technique is generally applied using data in the extreme ultraviolet (EUV) bands, joint modeling of EUV and high-frequency radio data has also been employed occasionally, providing more accurate estimates of plasma density \citep{landi2008}. However, this technique of deriving the plasma density has several limitations ranging from the availability of sufficient temperature coverage provided by broadband filters, inaccurate knowledge about the density of the relevant ionic species, other assumptions made during estimating DEM, and more importantly, the essentially unknown line-of-sight (LOS) depth and variations of the plasma parameters \citep{guennou2012a, guennou2012b}. White-light observations of the solar corona made by coronagraphs are sensitive to the light scattered by the electrons due to Thomson scattering and hence can be inverted to determine the off-limb coronal density \citep[e.g.][]{VanDeHulst1950, altschuler1972}. Accurate density estimates using a single LOS observation can only be made after assuming the geometry of the density structure. In most cases, the geometry is assumed to be spherically symmetric \citep{VanDeHulst1950, Hayes2001}. In extreme cases, it is possible that the geometry of the density structure is such that it remains completely invisible from one LOS, but is very bright from another \citep{vourlidas2000}. Spectral line observations, while they can provide accurate plasma diagnostics with reduced LOS effects, including density and temperature \citep{delzanna2018}, are generally limited to the solar disk and regions close to it. 

Low-frequency radio emission generally arises from a coronal height range that is not regularly covered by other density-sensitive probes. The Atmospheric Imaging Assembly \citep[AIA;][]{lemen2012}, onboard the Solar Dynamics Observatory \citep{pesnell2012}, which is widely used for estimating densities through the DEM analysis, has a FOV up to approximately 1.2 $R_\odot$. The widely used coronagraphs, detectors C2 and C3 of the Large Angle Spectroscopic Coronagraph \citep[LASCO;][]{Brueckner_1995} onboard the Solar and Heliospheric Observatory \citep[SOHO;][]{Domingo_1995} spacecraft, sample the corona beyond a heliocentric distance of about 2.2 $R_\odot$. Due to this, density models are generally used to estimate the coronal density between 1.2 and 2.2 $R_\odot$. In some instances, density models are also employed. Some widely used density models include the Newkirk model \citep{newkirk1961}, Saito model \citep{saito1977}, Leblanc model \citep{leblanc1998}, etc. However, the model-predicted density and the ``true" density can differ by up to an order of magnitude. Hence, developing new diagnostic techniques to probe coronal density, particularly in the observational gap between the coverage of coronagraphs and instruments like AIA, is crucial.

The Solar Ultraviolet Imager \citep[SUVI;][]{darnel2022} onboard the GOES instrument offers a large FOV (a square of sides around $3.2 R_\odot$)  and can significantly cover the gap between AIA and LASCO/C2, despite the fact that routine DEM analysis has been unavailable. Similarly, instruments like the K-coronagraph (K-Cor) of the COronal Solar Magnetism Observatory (COSMO) can also bridge this gap. Additionally, the methods used to estimate the density from these observations make their own set of assumptions, which can also reduce the robustness of the density estimates. Hence, it is also important to compare the density estimates obtained from different complementary sets of observations.

Low-frequency radio observations provide an independent means to determine the coronal density.  At the observed long wavelengths, the radio emission from the quiet solar corona predominantly arises from thermal free-free (Bremsstrahlung) emission, which is directly dependent on the temperature and density of the coronal plasma. In this work, we employ the technique developed in \citet{mercier2015}, henceforth referred to as M15, to determine the coronal density from radio observations. We compare these density estimates with those from popular empirical density models and those obtained by inverting the white-light coronagraph data. Additionally, we compare the density estimates obtained from these different observational techniques with those from a data-constrained magnetohydrodynamic (MHD) simulation, providing a comprehensive evaluation of the density diagnostics across different methodologies. 

Determining the coronal densities using the radio observations presented here requires flux density measurements at heliocentric distances beyond the unit optical depth surface, where the emission becomes fainter. Hence, a necessary requirement for obtaining the density estimates using this technique is to image the quiet sun at a high dynamic range.  While this is a challenge for most radio instruments, the newly commissioned Long Wavelength Array at the Owens Valley Radio Observatory (OVRO-LWA\footnote{An instrument paper is being prepared. See \citealt{Anderson2018} for a description of the instrument prior to the latest expansion.}), with its dense array configuration and 352 dipole antennas, is perfectly suitable of fulfilling this requirement on a routine basis. In this work, we use high dynamic range radio images from the OVRO-LWA to determine the coronal density from about 1.7--3.3$R_\odot$, for a few selected days. These radio-derived densities are then compared with estimates obtained from white-light observations using the polarized brightness (pB) images from the LASCO/C2 coronagraph. LASCO/C2 images the corona over a FOV extending from approximately $\sim2.2$--$6.5~R_\odot$, making it well-suited for measuring electron density along with those obtained from OVRO-LWA. Together, these instruments enable a comprehensive assessment of the coronal electron density spanning a broad range of heliocentric distances and physical regimes.

We summarize the methodology of density estimation in Sections \ref{sec:technique}, \ref{sec:MASmodel}, and \ref{sec:WLtechnique}. In Section \ref{sec:obs_analysis}, we briefly describe the data selection criteria and the analysis procedures. The results of the study are presented in Section \ref{sec:results}, followed by a broader discussion and conclusions in Section \ref{sec:discussion}.

\section{Density estimates using radio maps} \label{sec:technique}

In this work, we follow the density estimation method outlined in M15. Here we provide a brief overview of the key steps. The reader is referred to M15 for a detailed description.

The brightness temperature at any location away from the solar limb\footnote{Limb is defined as the heliocentric distance beyond which the optical depth becomes very small.} is given by 
\begin{equation}
T_B=\int_{ray \, path}T_e(\tau)d\tau 
\label{eq:brightness_temp_original}
\end{equation}

where $T_B$ is the observed brightness temperature, $T_e$ is the electron temperature, $\tau$ is the optical depth, and LOS is the line of sight. The optical depth is given by 
\begin{equation}
    \tau= \int_{ray \, path} \alpha dx, \quad \alpha=\frac{\zeta n_e^2}{\mu f^2T_e^{3/2}}
\end{equation}

$\alpha$ is the absorption coefficient of thermal plasma, $n_e$ and $T_e$ are the electron density and temperature of the plasma, $f$ is the observing frequency, $\zeta$ is a constant, equal to 0.23 in CGS units, and $\mu$ is the refractive index. 
Since the majority contribution of the optical depth comes from a small region close to the Sun, in comparison to the total distance between the Sun and the Earth (the gradient of density with heliocentric distance is very high), we can approximate equation \ref{eq:brightness_temp_original} as 
\begin{equation}
     T_B=T_e\tau
\end{equation}
Here, we have assumed that over the small region which contributes most to the integral in equation \ref{eq:brightness_temp_original} is isothermal.

Let us denote the point of closest heliocentric distance of the ray $r=r_1$. Assuming that, locally, the corona is in hydrostatic equilibrium, the primary contribution to the opacity comes from a thin layer of thickness $\sqrt{\pi H r_1}$, around $r=r_1$, where $H$ is the local hydrostatic scale height and is given by $H=k_B T_H/(\bar{m}g)$. $k_B$ is the Boltzmann constant, $T_H$ is the local hydrostatic temperature, $\bar{m}$ is the mean particle mass and is equal to 0.6$m_p$ for helium abundance of 0.1. $m_p$ is the proton mass, and $g$ is the local acceleration due to solar gravity. We ignore the density dependence of $\mu$ and assume that density only enters the equation through its dependence on $\alpha$. Under these assumptions, the relation between the observed brightness temperature and density is given by

\begin{equation}
    n_e(r)=\left(\frac{\bar{m}g(r)}{\pi k_B r} \right )^{0.25} \zeta^{-0.5} \left( \frac{T_e}{T_H(r)}\right)^{0.25} f T_B(r)^{0.5} \label{eq:density}
\end{equation}

However, there are several issues in directly using equation \ref{eq:density}. First of all, $T_H$ depends on the density distribution and hence is unknown a priori. Equation  \ref{eq:density} suggests that the brightness temperature at sky coordinate $r$ determines the density at $r$. However, this will only be valid if $\omega/\omega_p>>1$ throughout the ray path, where $\omega,\omega_p$ are the observing frequency and plasma frequency, respectively. In general, light ray will undergo refraction, and hence the ``true" point of closest approach of the ray will be different, say $r_{min}$. Thus, the density at location $r_{min}$ determines the observed brightness temperature at sky coordinate $r=r_1$. The deflection of the light ray is also dependent on the density and cannot be known a priori. 

To solve these issues, we take an iterative approach as outlined in M15. An improved value of $T_H$ can be obtained from the obtained density model, and then using the new improved estimate of $T_H$ to recalculate the density model in the next iteration. The method to take into account small deviations of the ray is also provided in M15. We describe the method briefly here. From Snell's law of refraction in spherical geometry $\mu r \sin i =\text{constant}$, where $i$ is the incidence angle on the spherical iso-density layers. Assuming that the angle of incidence does not change much, we obtain $r_1/r_{min}=\mu(r=r_{min})/\mu(r=r_1)$. Since most of the opacity along the ray arises from $r=r_{min}$, this also implies that $\mu(r=r_1)\approx 1$. Thus, we obtain the equation that is relevant for iteration.
\begin{equation}
    \frac{r_1}{r_{min}}=\mu(r=r_{min})
\end{equation}

We start the iteration assuming that $T_H=1.4$MK and that the ray suffers no refraction. Using this assumption, we determine the density model. We use this derived density model to determine a new estimate of $T_H$. We also use the density model to determine $r_{min}$ and assign the obtained density estimate at $r=r_1$ in the sky plane to $r=r_{min}$. We iterate using these new parameter estimates until the density model obtained from this procedure is consistent with the one used to determine $T_H$. In practice, we stop the iteration when the mean relative difference between the density estimates at a fixed heliocentric distance is smaller than $5\%$. In our analysis, we have assumed that $T_e$ is 1 MK. However, we have also verified that assuming a temperature of 0.8 MK does not significantly affect the results. This weak dependence on temperature is also expected, as density depends only on $T_e^{0.25}$, as compared to a much stronger dependence on brightness temperature and observation frequency.

\section{MAS model}\label{sec:MASmodel}

The Magnetohydrodynamic Algorithm outside a Sphere (MAS) is a global, data-driven MHD model developed by Predictive Science Inc. that simulates the solar corona by solving the full set of time-steady MHD equations in spherical geometry \citep[e.g.][]{mikic1996,mikic1999, lionello2009, riley2011}. To derive the coronal electron density, the MAS model takes as input synoptic magnetograms—typically from instruments like SDO/HMI—to specify the photospheric radial magnetic field. Using these boundary conditions, the model extrapolates the magnetic field into the corona while self-consistently evolving plasma parameters including density, velocity, pressure, and temperature. The predicted distribution of the plasma parameters is available publicly at \url{https://www.predsci.com/mhdweb/data_access.php}. For this work, we used the MAS model corresponding to Carrington Rotation 2283, which provides global, three-dimensional electron density distributions over the full solar corona. The model output includes radial density profiles at different latitudes and longitudes. We took an average over all latitudes and longitudes to derive the radial dependence of density, which we compared with the density profile derived using other techniques.

\section{Density estimates using White-light Coronagraph} \label{sec:WLtechnique}

The electron density in the solar corona can also be estimated using white-light observations from the ground and space-based coronagraphs. These observations are primarily the result of photospheric light scattered by free electrons in the corona, through the Thomson scattering process \citep{Billings1966}, and by interplanetary dust in orbit about the Sun. The electron-scattered component, known as the K-corona, is polarized, whereas the dust-scattered component, known as the F-corona, is generally considered to be unpolarized in the height regime we are considering. By analyzing the polarized brightness (pB) or the total brightness components of this scattered light, one can infer the spatial distribution of electron density.

Here, we calculate the electron density using the classical inversion technique originally described by \cite{VanDeHulst1950} and further developed in \cite{Hayes2001} based on pB observations. This method involves inverting the line-of-sight (LOS) integrated pB brightness to retrieve the electron density as a function of radial distance along a given position angle (PA), under the assumption of spherical symmetry around the Sun. The observed pB intensity in a given white-light image is the LOS integral of the local electron density weighted by the Thomson scattering function. The coronal electron density $n_e(r)$ as a function of the radial height from the Sun's center $r_s$ is determined by the inversion of the following equation \citep{VanDeHulst1950, Billings1966, Hayes2001}.

\begin{equation}
pB(\rho) = C\int_\rho^\infty n_e(r_s) [A(r_s)-B(r_s)] \frac{\rho^2}{r_s\sqrt{(r_s^2-\rho^2)}} dr_s
\label{eq:pb_ne_relation}
\end{equation}

where $C$ is the conversion factor, $A(r)$ and $B(r)$ are the geometrical scattering factors \citep{VanDeHulst1950, Billings1966}, and $\rho$ is the projected heliocentric distance (i.e., the impact distance) of a specific point, in the plane of the sky, through which LOS passes. $dr_s$ is an infinitesimal element in heliocentric radial distance along the LOS, related to distance along the LOS $y$ by $r_s = \sqrt{\rho^2 + y^2}$. Although the integration is carried out along the LOS, it is expressed in terms of radial distance $r_s$ and $dr_s$.

\begin{figure}
\centering
    \includegraphics[scale=0.1]{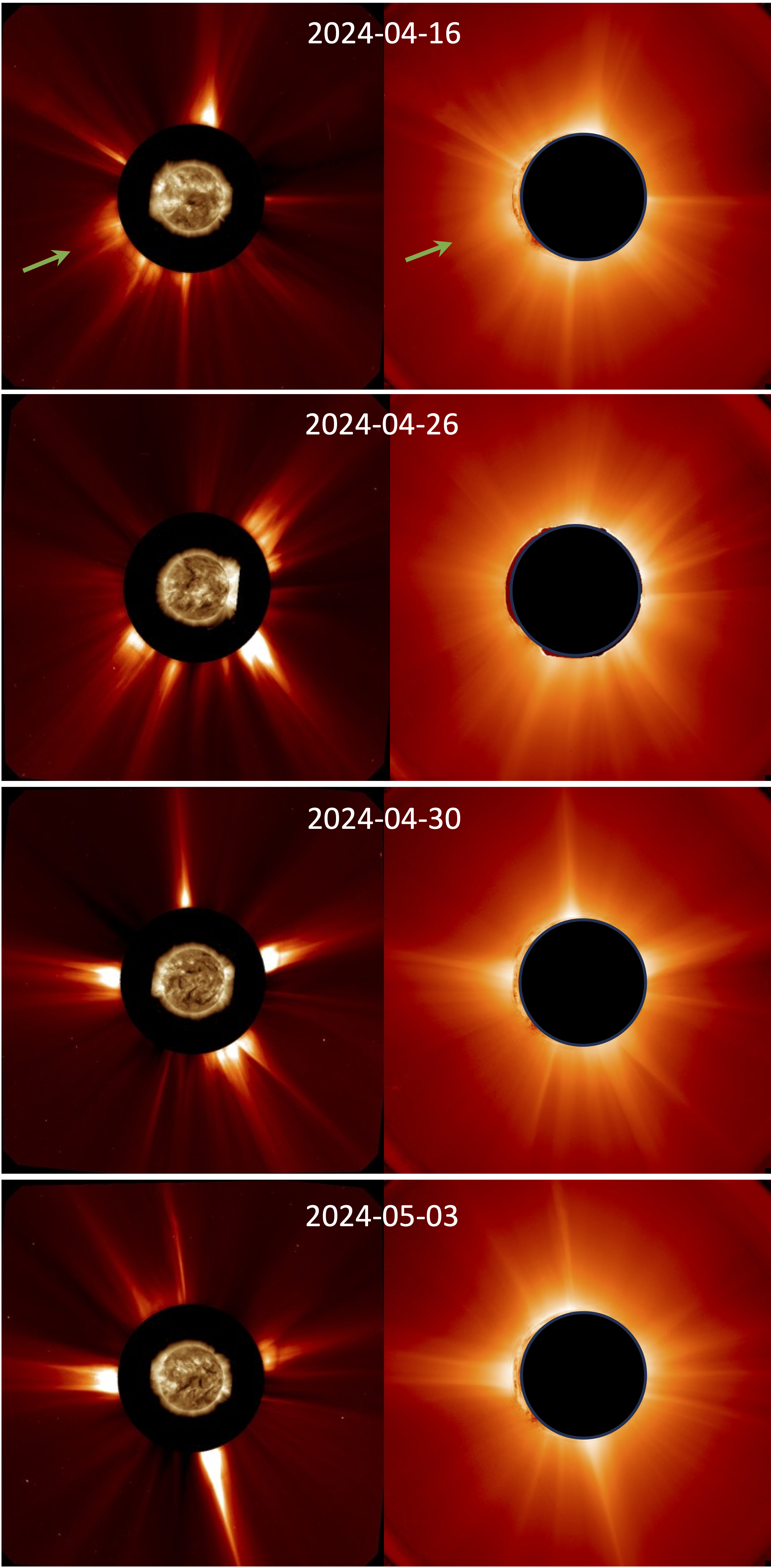}
    \caption{White-light total brightness (left column panels) and polarized brightness (right column panels) images from  LASCO/C2 coronagraph observations for all the dates under study. The total brightness images are composited with extreme ultraviolet images at  $193$~{\AA} from SDO/AIA.}
    \label{fig:c2_img}
\end{figure}

To ensure the accuracy of the quiescent coronal background density measurements, the pB images used for each date in our analysis were carefully selected to be free from any significant coronal mass ejection (CME) activity, which carries large concentrations of mass that can significantly distort the brightness and polarization of the white-light emission, thereby introducing errors in the derived densities of the undisturbed background corona. The corresponding LASCO/C2 images for each date, showing coronal features in total (left column panels) and polarized brightness (right column panels), are displayed in Figure~\ref{fig:c2_img}.

The total brightness images capture combined contributions from both the electron-scattered K-corona, the dust-scattered F-corona, galactic starlight, and instrumental stray light \citep{Billings1966, Kimura_1998}. LASCO/C2 images have been calibrated to remove the known small stray light component; however, a recently identified stray light contribution remains uncorrected in the current data processing (K. Battams' private communication). In contrast, the pB images provide only the polarized component of the K-corona, effectively suppressing the unpolarized F-corona and any residual stray light. As a result, some features visible in total brightness are not present in pB images, highlighting the electron density structures of the corona, as seen in Figure~\ref{fig:c2_img}. For example, the streamer region over the southeastern limb, marked by green arrows in Figure~\ref{fig:c2_img}, is clearly visible in the total brightness image but appears much fainter in the corresponding pB images, indicating that the streamer is away from the plane of the sky. Note that the total brightness and pB images are byte-scaled independently to increase the contrast and improve the visibility of features in each image.

The density estimates from pB inversion are derived at every 3-degree interval in position angle, covering the full 360-degree solar disk, allowing for full angular coverage of the corona. At each PA, pB data are sampled radially every $0.024~R_\odot$ from $2.2$ to $6.5~R_\odot$ of LASCO/C2 FOV. For each PA, density radial profiles resulting from poor fits, likely due to intensity artifacts or a low signal-to-noise ratio, are systematically flagged and excluded from the resulting overall density. Although the density profiles exhibit noticeable variations across polar and equatorial regions of the corona, we compute an average of all the valid profiles, consistent with the approach followed in radio-based estimates. The white light profiles are plotted only up to 3.5~$R_\odot$, corresponding to the heliocentric distance up to which emission is detected in the multi-frequency radio images.

\section{Data analysis} \label{sec:obs_analysis}

 All data products used in this study are publicly available from the OVRO-LWA database\footnote{https://www.ovsa.njit.edu/lwadata-query}. 
 These data products are produced using the OVRO-LWA solar data analysis pipeline, which will be described in detail in Mondal et al. (in preparation). Below, we briefly describe the procedure used for the data pipeline.

In the data analysis pipeline, we correct for both instrumental effects as well as effects due to the propagation of radio waves through the ionosphere. OVRO-LWA is a very stable instrument due to its simple architecture. Hence, instrumental effects are typically calibrated a few times a month. For this purpose, we generally use nighttime data, when bright astronomical sources such as Cygnus A, Cassiopeia A, Taurus A, or Virgo A are visible in the sky. Any variation in the antenna gains is corrected using self-calibration techniques. We then apply the antenna gains derived from night-time calibration data, as well as those obtained through self-calibration, to obtain the final calibrated data. Next, we subtract bright astronomical sources present in the sky and change the phase center to the solar location. We use these data to produce images with a FOV of $96\ R_\odot$\footnote{A square region, centered around the location of the Sun, of size around $24 R_\odot \times 24 R_\odot$ is cropped from these images and archived. These archived images have been used for this work.} at both 384 kHz and 5 MHz frequency resolution. While OVRO-LWA records data between 13--87 MHz, due to the presence of strong radio frequency interference at low radio frequencies, the pipeline does not produce imaging data products below 32 MHz. { The solar images, thus produced, can have a location different from the true sky location due to ionospheric effects. This is caused by the gradient of ionospheric total electron count (TEC), and is expected to have a $f^{-2}$ dependence, where $f$ is the observation frequency. We use the known frequency dependence of the ionospheric shifts to determine the TEC gradient using the multi-frequency images available from OVRO-LWA and correct for it. This procedure ensures that the solar center after this correction matches the true location of the Sun in the sky plane.}

\begin{figure*}
\centering
    \includegraphics[scale=0.75]{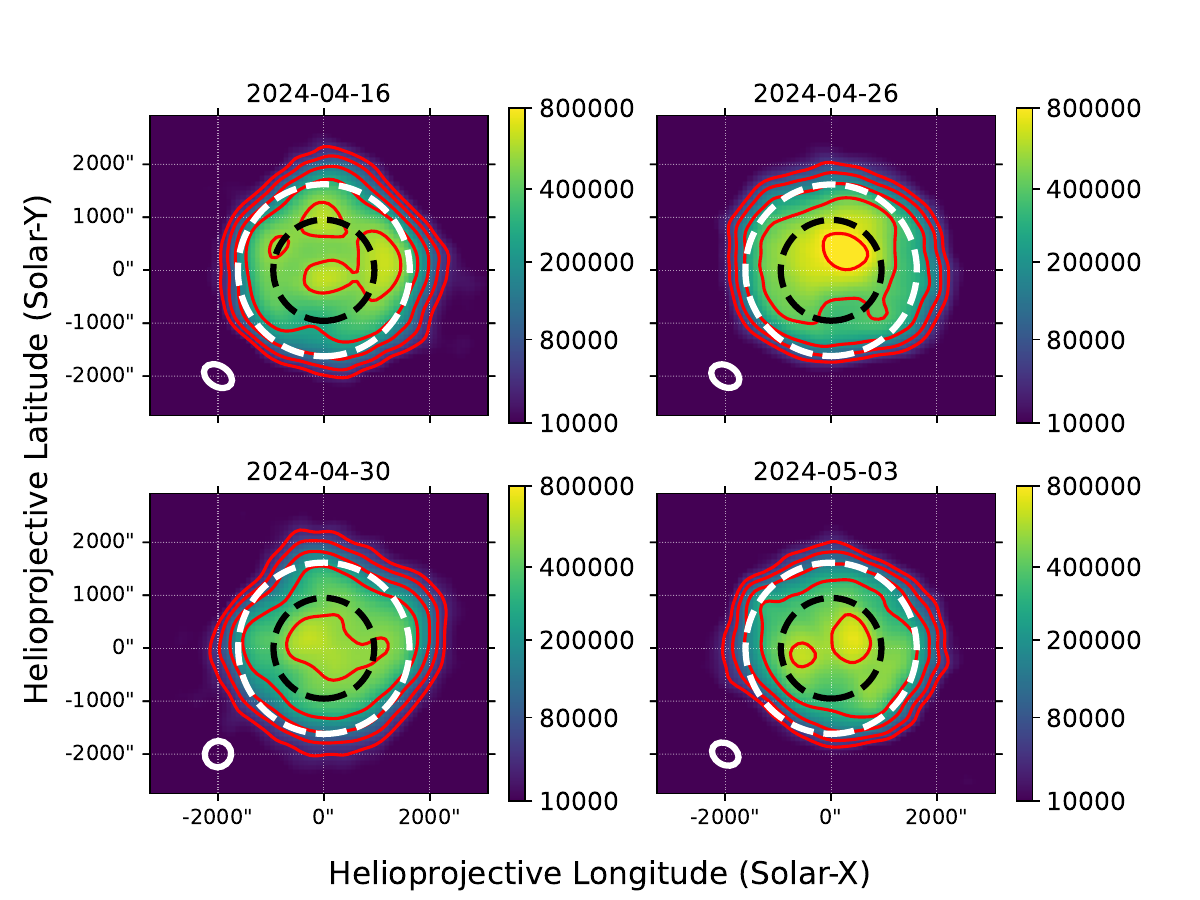}
    \caption{Example radio images at 71 MHz for different days. Colors indicate the brightness temperature. The lowest contour level is at $5\%$ of the peak and then increases in multiples of 2. The black dashed circle represents the optical disk of the Sun.  The white dashed circle indicates the heliocentric distance beyond which we apply the density estimation technique presented here, for this frequency. The white ellipse drawn at the bottom-left corner of each panel shows the corresponding instrumental resolution.}
    \label{fig:example_solar_imgs}
\end{figure*}

For the purpose of this work, we chose quiet Sun data from four days in 2024, April 16, April 26, April 30, and May 3. Radio images at an example frequency of 71 MHz are shown for four chosen times in Figure \ref{fig:example_solar_imgs}. The lowest contour level is at $5\%$ of the peak intensity and then increases in multiples of 2. The optical solar limb is shown with a black dashed circle. The high dynamic range of these images is evident, and images at other frequencies on these dates are also comparable. 


\section{Results} \label{sec:results}

\begin{figure*}
\centering
\includegraphics[scale=0.5]{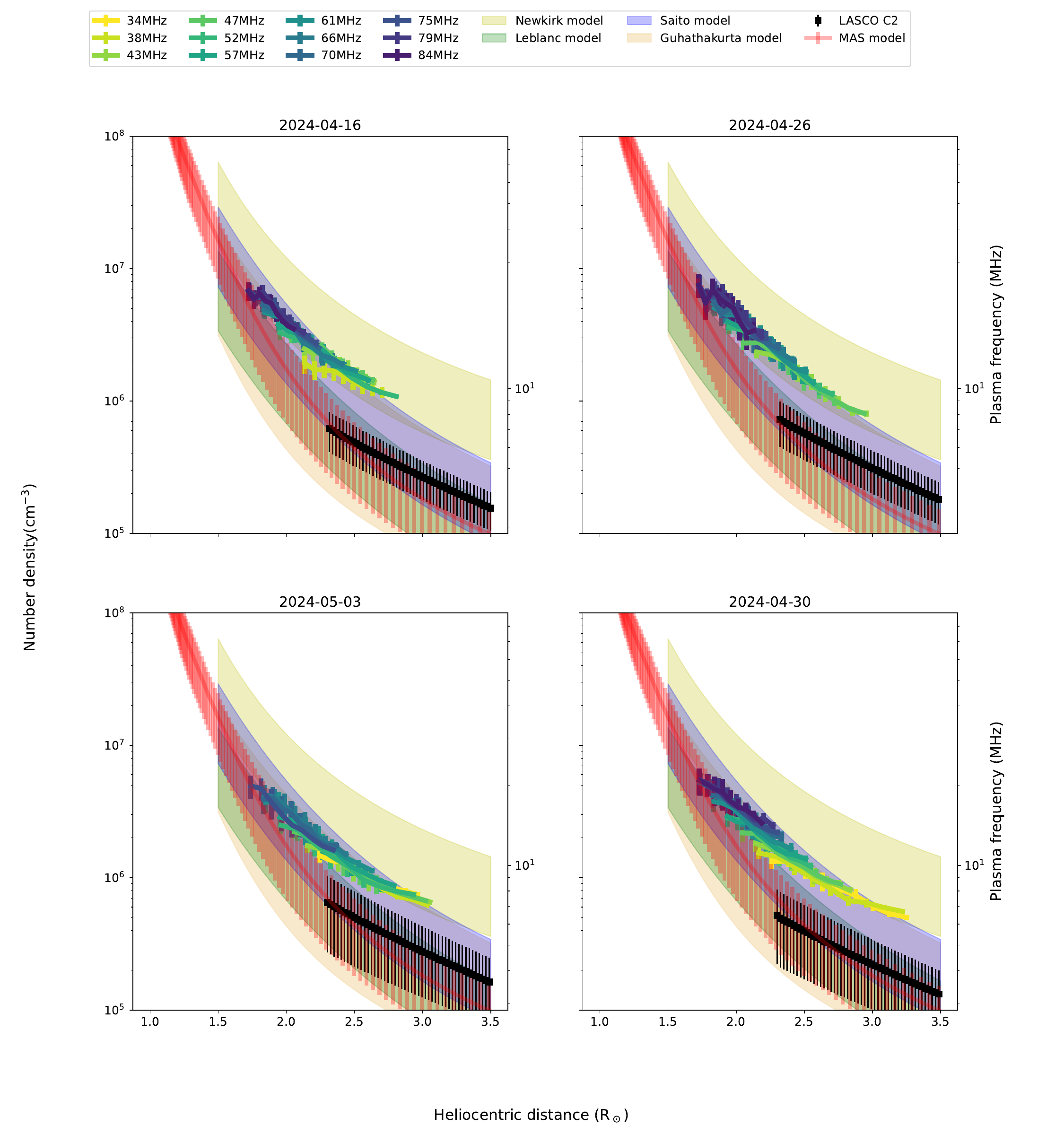}
\caption{Density estimates obtained on different dates are shown in separate panels. The date corresponding to each panel is indicated in the respective title. The comparison highlights the density estimates from our dataset, based on radio and white-light observations, consistent with previous findings and model-predicted densities.}
\label{fig:density}
\end{figure*}

In Figure \ref{fig:density}, we show the density estimates obtained from this work. The radial density profiles have been obtained by averaging the derived densities over all position angles. The error bar represents the standard deviation of the densities in each radial bin. Each radial bin is of size $0.1R_\odot$. The black solid lines show the densities estimated from the white-light observations. These densities were estimated using the nearest available pB white light image from the LASCO/C2. The radio-derived densities based on OVRO-LWA data are shown using different colored solid lines. The colors vary smoothly from dark blue to yellow as the frequency of the observation decreases from 84 to 34 MHz. The error bars shown for each radio-derived density are a measure of the random uncertainty associated with the measurement.
We have also shown three widely used density models, namely the Newkirk model \citep[yellow,][]{newkirk1961}, the Saito model \citep[blue,][]{saito1977}, the Leblanc model \citep[green,][]{leblanc1998} and the Guhathakurta model \citep[wheat,][]{guhathakurta1996}. The bands represent the densities, considering that the actual density can vary between 0.5--2 times the value predicted by the Newkirk, Saito and the Leblanc models. { For the Guhathakurta model, the band represents the variation betweewn the polar and equatorial density model.} Although this range may indeed appear large, scaling factors of the order of 6 \citep[e.g.][]{mccauley2018, zhang2023, kumari2023} are commonly applied to the density models by solar radio astronomers to make radio observations consistent with other observables. For comparison, we have also shown the radial density profile obtained from the MAS model using a red solid line. {The errorbars correspond to the standard deviation of the density values at a fixed radius.} The MAS model corresponds to the same Carrington map as the observations presented in this study.

We note the following inferences from Figure \ref{fig:density}.
\begin{enumerate}
    \item The density estimates obtained from the radio and the white-light observations are consistent within a factor of $\approx 2$. 
    \item While the Saito density model seems to be quite consistent with both white-light and radio estimates, the Newkirk model overestimates the densities compared to the white-light results. On the other hand, the Leblanc model slightly underestimates the density. 
    \item At large heliocentric distances, the density estimates of the MAS model are more consistent with the white-light estimates compared to the radio-derived values. However, the discrepancy between the MAS model and radio data decreases significantly with a decrease in the radial distance and is approximately consistent with the radio estimates at radial distances smaller than $\approx 1.7 R_\odot$.
\end{enumerate}

\begin{figure}
    \centering
    \includegraphics[width=\linewidth]{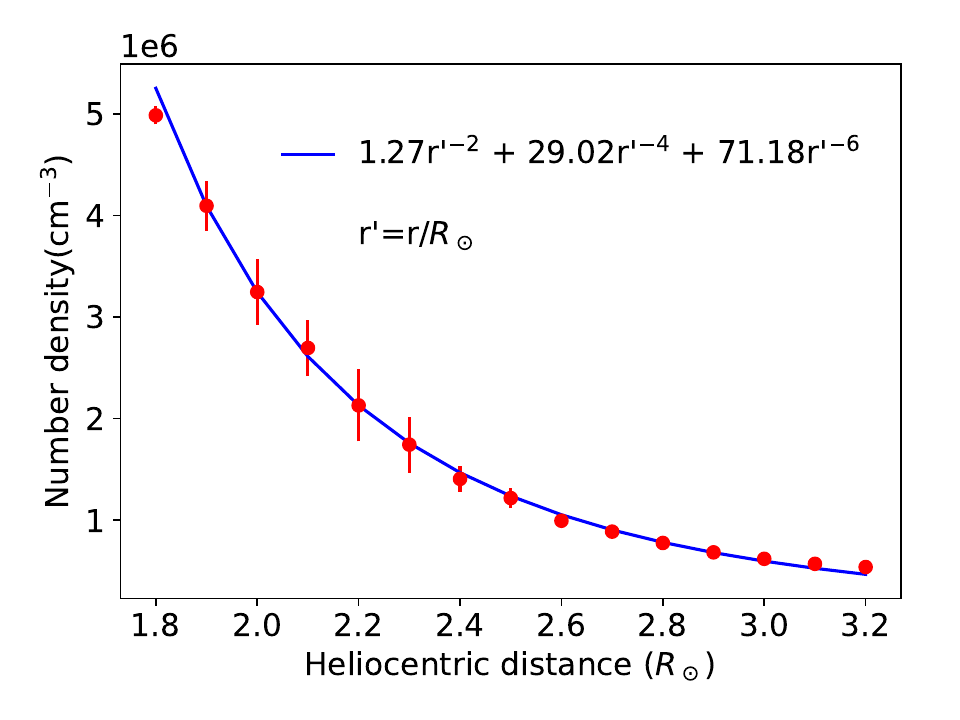}
    \caption{Red points show the densities derived using OVRO-LWA data from April 30, 2024. The blue line shows the fitted model. The fitted coronal density model is provided in the figure as well.}
    \label{fig:density_model}
\end{figure}

In Figure \ref{fig:density_model}, we use the obtained density estimates from April 30, 2024, to generate a coronal density model for that day. { The error bars correspond to the standard deviation of the spherically averaged density values with frequency\footnote{This leads to the measurement artifact that the error bars are negligible at higher coronal heights.}.} We choose to use this particular day because the density estimates for this day cover the largest heliocentric distance and are best suited for generating a coronal density model. While we have used the functional form of the Leblanc density model, the coefficients of the model are obtained using radio-derived densities. The obtained coronal density model is given by
\begin{equation}
\begin{aligned}
    \frac{n_e(r)}{10^6\ \mathrm{cm}^{-3}}=&1.27\left (\frac{r}{R_\odot}\right )^{-2} + 29.02\left (\frac{r}{R_\odot}\right )^{-4} + \\
    & 71.18\left (\frac{r}{R_\odot}\right )^{-6}
\end{aligned}
\end{equation}

\section{Discussion} \label{sec:discussion}

Coronal densities beyond $\sim 1.2-1.3 R_\odot$ are generally estimated using white-light observations. Among the few routinely available diagnostics for this purpose, total brightness and polarized brightness observations are available from LASCO/C2. However, estimating the density profiles from these observations requires several underlying assumptions, which may impact the accuracy of the estimated densities. Key assumptions include the following:

\begin{itemize}

\item {From Equation \ref{eq:pb_ne_relation}, it is evident that we assumed that the density along the LOS depends primarily on the radial distance from the solar center. Although we do not enforce the same radial density variation across the sky plane, as each projected location on the sky plane is inverted independently, the density obtained at each point effectively represents a LOS-integrated quantity, weighted by the scattering efficiency. Thus, the derived density at each projected location on the sky plane can be interpreted as a weighted average over the LOS, dominated by contributions near the plane of the sky.}

\item {Plane of sky bias: An advantage of using pB measurements is that the Thomson scattering efficiency is concentrated closer to the point along the LOS that is perpendicular to the line from the Sun center, where it is maximum. The scattering efficiency factor falls to approximately $50\%$ of the maximum value at approximately $\pm30^{\circ}$ from the plane of the sky -- the plane perpendicular to the LOS through the Sun's center \citep{Michels_1997}. The Thomson scattering, along with the degree of polarization of the scattered light, peaks when the electrons lie in the plane of the sky. As a result, the majority of the pB emission observed is favored with the structures that are closer to this plane, which provides the inversion method with high fidelity in that region. Conversely, contributions from electrons located significantly far from the plane of the sky are substantially weaker.} 
\end{itemize}

While the LASCO/C2 FOV covers the region where the corona transitions from a structured, close-field configuration to more open, solar-wind dominated conditions, enabling the characterization of the extended corona and the associated density gradients, it lacks coverage below approximately 2$R_\odot$. Consequently, density variations in the inner corona, closer to the Sun below 2$R_\odot$, remain inaccessible to the current white-light coronagraphs. This region, however, is probed by radio observations, offering complementary diagnostics for the near-Sun coronal environment, as addressed in this study. 
Here, we utilized the high dynamic range radio images available from the OVRO-LWA to estimate the density of the solar corona over the heliocentric distance of approximately $1.7-3.5 R_\odot$. The key assumptions underlying the density estimation from the radio imaging are outlined below. We have assumed that:

\begin{itemize}
\item The dominant emission mechanism is thermal bremsstrahlung, 
    \item the dominant contribution to the radio opacity originates from a localized region near the point along the ray path where the density reaches its maximum, 
    \item the local density variation can be described by a hydrostatic equilibrium distribution, close to this maximum density point, and 
    \item the ray suffers only a small deviation from its straight-line trajectory, which implies that the local plasma frequency is significantly smaller than the frequency of the observation.
\end{itemize}

It is evident that the assumptions made when estimating the densities using white light and radio observations differ significantly. For example, in white-light observations, most of the contribution arises from density structures close to the plane of sky, and radio observations are affected equally by all the density structures within the ray path. However, equation \ref{eq:density} implicitly assumes that the majority of the contribution to the opacity arises from a location close to the sky plane. Additionally, refraction along the ray path can alter the ray direction, which affects the obtained estimates from the radio data. From Figure \ref{fig:density}, it is clear that the inferred plasma frequency is about 3--4 times smaller than the observation frequency. While our efforts to account for the deviation of the ray path from a straight-line trajectory, as well as the low plasma frequency compared to the observation frequency, minimize this effect, it will still be present and will affect our results. However, it is hard to quantify the effect of this issue on the density estimates generated here. Similarly, without knowing the density structures along the LOS, it is difficult to quantify the systematic uncertainties associated with the estimates obtained from the white-light data. 

In this work, we have also compared the widely used coronal density model with the densities obtained by inverting the radio and white-light data.  We find that, while the Saito density model seems to be quite consistent with both white-light and radio estimates, the Newkirk model overestimates the densities compared to the white-light results. On the other hand, the Leblanc model slightly underestimates the density.  This is not very surprising, as these models were themselves derived using multiple physically and phenomenologically motivated assumptions. Additionally, the coronal conditions of the data used in deriving these models can also be very different than the data used here. For example, \citet{newkirk1961} mentioned that the densities they derive during a particular sunspot maximum are twice that reported by \citet{VanDeHulst1950}, and hence, while the functional form of the Newkirk model is the same as that used in \citet{VanDeHulst1950}, the electron densities have been scaled by a factor of 2.2. The Saito density model, however, was derived using data from solar minimum conditions. \citet{saito1977} specifically chose days during which the quiet coronal density was not significantly influenced by coronal holes and streamers. In this work, we have also reduced the effect of heliospheric structures and coronal transients by averaging across position angles. This may explain why the Saito model is consistent with the radio data.  \citet{leblanc1998} notes that their model may not be applicable if the active regions undergo changes within the time period of approximately 4 days needed for the solar wind to reach 1 AU from the Sun. During solar maximum, such changes are quite common, and hence, the scaling factor obtained using the coefficients of the Leblanc model may be inaccurate.

\begin{figure*}
    \centering
    \includegraphics[width=\linewidth]{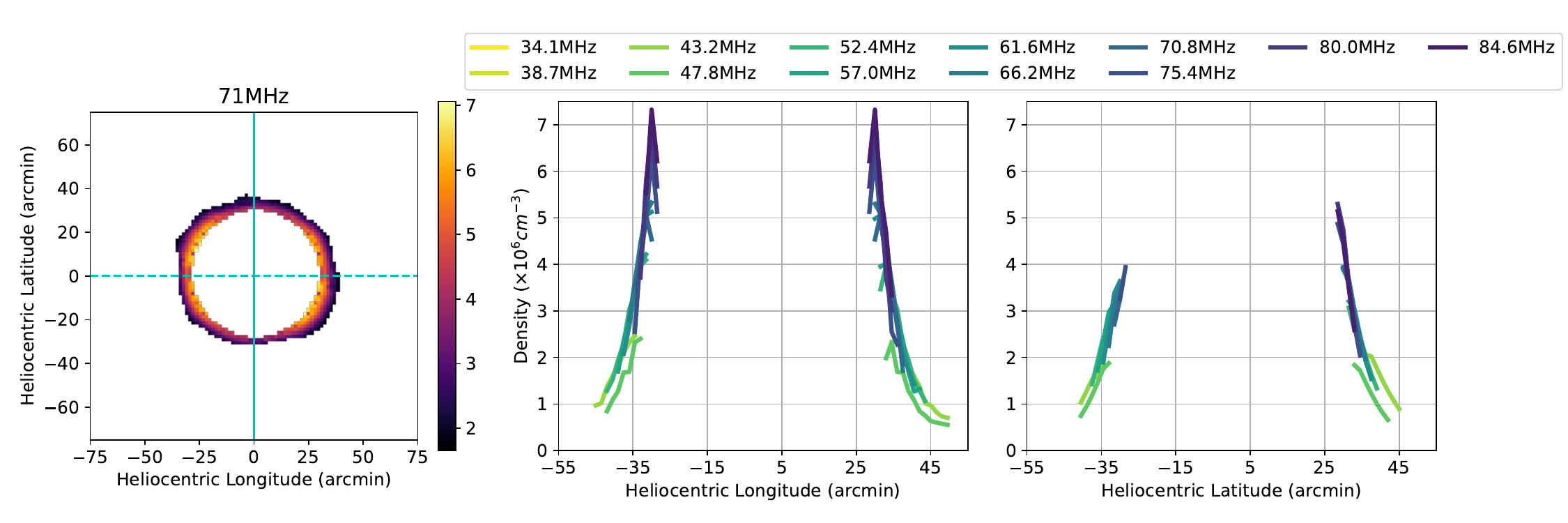}
    \caption{Left panel: Shows the densities obtained by inverting solar maps at 71 MHz from April 26, 2024. Middle and right panel: Shows the density variation along 1D cuts through the center of the Sun in the sky plane along the equator and poles, respectively. }
    \label{fig:1D_density_cut}
\end{figure*}

In this work, we have averaged across all position angles to determine a spherically symmetric density model. However, observations at other wavebands suggest that the coronal densities are dependent on the position angle. In particular, the density at the poles is lower than that at the equator. From the images shown in Figure \ref{fig:example_solar_imgs}, it is clear that the angular extent of the solar disc along the equator and the poles is different. \citep{zhang2022_quiet_sun_density} reported a difference in the size of the solar disk of about $400\arcsec$ between the equatorial and polar radii at around 78 MHz. The OVRO-LWA has sufficient angular resolution to measure this difference. Hence, it can be expected that it should also be able to measure the density difference between the polar and equatorial regions. In Figure \ref{fig:1D_density_cut}, we show the density map as well as density cuts along the equator and poles for April 26, 2024. We find that data from each frequency predicts a slightly different density, and hence their variation can be treated as a measure of uncertainty. We also note that the densities obtained both at the poles and the equator are consistent if we consider the variation across frequencies as a measure of uncertainty. This also justifies the choice of averaging across position angles, as the density estimates we report here are not suitable for measuring the density difference at varying position angles.

Equation \ref{eq:density}, which is used to obtain the radio-derived density estimates, assumes that the heliocentric distance of any point is much smaller than the local density scale height. We find that the density scale height obtained from the inverted density profile is approximately $0.6R_\odot$. Considering that we have used the radio data to determine densities between $1.7-3.5 R_\odot$, the ratio $H/r$ varies between $\sim 0.3-0.17$. Thus, at the lower coronal heights, the approximation is not strictly valid, and hence the inverted densities can have some systematic uncertainties.

{ Here, we have used data near the peak of solar cycle 25. During this time, the Sun was quite active, and hence times devoid of activity were chosen with care such that this formalism could be applied. Additionally, during this time, the Sun is expected to have complex coronal structures, which limits the accuracy of the radial density profile assumption employed in this work.} However, despite these issues, we find that the density estimates obtained from these different techniques are consistent within a factor of two over the entire heliocentric distance common to the radio and white-light images. We also note that the radio-derived density estimates are similar to those obtained from the MAS model over the entire heliocentric distance spanned by the radio estimates. {This formalism is expected to work even more robustly than shown here during solar minimum conditions, when the coronal structures are less complex, and the solar activity is significantly less.}
This suggests that observations with the OVRO-LWA can be used to regularly obtain coronal density estimates of the quiescent solar corona over the heliocentric distance of $\sim 1.7-3.5 R_\odot$, and can play a complementary role to the estimates obtained with the white light data.


\begin{acknowledgments}
The OVRO-LWA expansion project was supported by NSF under grant AST-1828784. OVRO-LWA operations for solar and space weather sciences are supported by NSF under grant AGS-2436999. SOHO is a joint mission of the European Space Agency (ESA) and the US National Aeronautics and Space Administration (NASA). LASCO was built by a consortium of the Naval Research Laboratory, USA, the Laboratoire d’Astronomie Spatiale in Marseille, France, the Max Planck Institute für Aeronomie in Lindau, Germany, and the School of Physics and Astronomy, the University of Birmingham, UK. S. S. and R. H. thank K. Battams, the principal investigator of LASCO, for all the valuable discussions on the LASCO/C2 calibration. P. Z. acknowledges support for this research by the NASA Living with a Star Jack Eddy Postdoctoral Fellowship Program, administered by UCAR’s Cooperative Programs for the Advancement of Earth System Science (CPAESS) under award 80NSSC22M0097.  A portion of this research was carried out at the Jet Propulsion Laboratory, California Institute of Technology, under a contract with the National Aeronautics and
Space Administration (80NM0018D0004).
\end{acknowledgments}

\bibliography{bibliography}
\bibliographystyle{aasjournal}



\end{document}